# Orbit Determination of Close Binary Systems using Lucky Imaging

F. M. Rica[1,2*], R. Barrena[3,4], G. Vázquez[5,7], J. A. Henríquez[6,7], and F. Hernández[7]

[1] *Astronomical Society of Mérida, C/Toledo, nº 1 Bajo, Mérida, , E-06800, Spain*
[2] *Double Star Section of LIADA, Avda. Almirante Brown 5100 (Costanera), S3000ZAA Santa Fe, Argentina*
[3] *Instituto de Astrofísica de Canarias, C/ Vía Láctea, s/n, La laguna, E-38205, Spain*
[4] *Depto. de Astrofísica, Univ. de La Laguna, C/ Astrofísico Francisco Sánchez, s/n, La Laguna, E-38201, Spain*
[5] *Astronomical Observatory MPC J-48, C/San Pancracio, nº 26, Urb. Mackay, San Cristóbal de la Laguna, E-38205, Spain*
[6] *Astronomical Society of Isla de la Palma, C/ A no, nº 2, Breña Baja, E-38712 , Spain*
[7] *Grupo de Observadores Astronómicos de Tenerife (GOAT), Spain*

**ABSTRACT**

We present relative positions of visual binaries observed during 2009 with the FastCam "lucky-imaging" camera at the 1.5-m Carlos Sánchez Telescope (TCS) at the observatorio del Teide. We obtained 424 CCD observations (averaged in 198 mean relative positions) of 157 binaries with angular separations in the range 0.14-15.40", with a median separation of 0.51". For a given system, each CCD image represents the sum of the best 10-25% images from 1000-5000 short-exposure frames. Derived internal errors were 7 mas in $\rho$ and 1.2º (9 mas) in $\theta$. When comparing to systems with very well-known orbits, we find that the rms deviation in $\rho$ residuals is 23 mas, while the rms deviation in $\theta$ residuals is 0.73º/$\rho$. We confirmed 18 Hipparcos binaries and we report new companions to BVD 36 A and J 621 B. For binaries with preliminary orbital parameters, the relative radial velocity was estimated as well. We also present four new revised orbits computed for LDS 873, BU 627 A-BC, BU 628 and HO 197 AB. This work is the first results on visual binaries using the FastCam lucky-imaging camera.

# 1 INTRODUCTION

The study of visual binary stars is long recognized as a basic key to the understanding of stellar structure, formation and evolution. Binary and multiple stars are prime targets for determining and calibrating basic stellar physics in general. The main importance in its study resides in stellar mass determination and in the mass-luminosity calibration. The stellar mass is determined from orbital information which can be obtained using a set of astrometric measures.

To determine the stellar masses accurately, high-quality orbital parameters and trigonometrical parallaxes are necessary. Hipparcos parallaxes made a significant progress in the orbital solutions. In several years, GAIA, the ESA astrometry space mission, will improve the distances measures. These highly accurate distances will allow astrophysics to enormously improve the precision in stellar masses determination. Most of the known orbital solutions for visual binaries are of marginal or poor quality, so it is important to obtain high-quality measures to improve their orbital parameters and obtain good stellar masses in combination with GAIA distances.

To date, new observational techniques can be used to measure binary systems with very small separation and larger dynamic ranges. Lucky imaging (see, e.g., Tubbs et al. 2003; Law, Mackay & Baldwin 2006) is one of the most advanced techniques to be applied to close binary stars. Same recent works: Daemgen et al. 2009; Lodieu, Zapatero Osorio, & Martín 2009; Bergfors et al. 2010; Schnupp et al. 2010; Labadie et al. 2011. Lucky imaging technique use CCD cameras capable of taking many images with exposure times short enough to reduce or freeze the changes in the turbulence of the Earth's atmosphere. Only the frames less affected by the atmosphere are shifted and stacked to form a single high-resolution image.

---

* Email: frica0@gmail.com

The work presented here details the first results of a binary star observation project using the new high-resolution imaging camera FastCam (Oscoz et al. 2008; Rodríguez Ramon et al. 2008) mounted to the 1.5-m Carlos Sánchez Telescope (hereafter TCS) at Teide Observatory. A sample of visual binaries was selected due to their need to be measured. Basically, our selection consists of neglected binaries or those with orbital solutions showing important residuals. The high-quality measures presented here provide necessary data to make significant progress with orbital solutions.

The Washington Double Star Catalog (Mason et al. 2001; hereafter WDS) database was used extensively in this work. The WDS, maintained by the United States Naval Observatory (USNO), is the most important database for astrometric double and multiple star information. A large amount of the historical measures for the binaries studied here were performed using micrometric eyepieces. This technique was the traditional method used during more than 2 centuries. Nowadays, measures using digital techniques such as speckle (e.g. Docobo et al. 2010; Orlov et al. 2010; Horch et al. 2011), adaptative optics (e.g. Liu, Dupuy & Leggett 2010), lucky imaging, etc. are more accurate. So it is important to increase the quality of the measures of each binary star. In this work, we have measured many binaries using digital techniques for the first time. Historical astrometric data were kindly supplied by Brian. D. Mason.

For several binaries, consistent residual trends demonstrated systematic runoff from published orbits. For some of these binaries, we were able to obtain new orbital elements that, although still preliminary, allow a better ephemeris estimate. We present new orbital parameters and other data (residuals, ephemerides, masses, parallaxes, apparent orbits, etc.) for 4 visual binary systems: WDS01032+2006 (LDS 873), WDS16492+4559 (BU 627 A-BC), WDS17184+3240 (BU 628), WDS23114+3813 (HO 197 AB). These are the first orbital parameters published for BU627 A-BC. Initial orbits were computed using the method developed by Docobo (1985) and later improved upon using the differential correction method presented by Heintz (1978a). The dynamical parallaxes were estimated using the Baize-Romaní algorithm (see Baize-Romaní 1946). All of the orbits presented here have been previously announced in the Information Circular of IAU Commission 26 (hereafter IAUDS).

The physical relation for the wide components of WDS16492+4559 were studied using optica BVI and infrared JHK photometry, as well as historical astrometric and kinematical data. Astrophysical properties for these components will also be determined and discussed.

On the other hand, because of their small binding energies, the importance of studies of wide binaries lies in that they are good sensors for detecting unknown mass concentrations that they may encounter along their galactic trajectories. For this reason, wide binary star systems have became objects of considerable theoretical and observational interest (e.g. Retterer & King 1982; Dommanget 1984; Lathan et al. 1984; Halbwachs 1988; Close & Richer 1990; Poveda & Allen 2004; Sesar, Ivezic & Juric 2008). They are relevant for understanding the processes of formation and dynamical evolution of our Galaxy. Thus, the present-day distribution of wide binaries can provide information about the disruption process as well as binary formation. Also, it is important to determine the characteristics and the frequency of binary stars in different stellar populations and evolutionary stages. In order to reach this aim, the astrophysical information such as magnitudes, colours, spectral types and velocities is fundamentally needed.

The organization of this paper is as follows. In § 2 we present the instrumentation and software used in this work. In § 3 we detail the composition of the observing list and the process of calibraton. In § 4 we inform about the method of orbital calculation. In § 5 we report our observational results (astrometric measurements and orbit calculations, new discoveries, radial velocities calculations and measurement capabilities). In § 6 we describe same systems (orbital binaries and wide components). And finally, in § 7 we present our conclusions.

## 2 INSTRUMENTATION AND DATA PROCESSING

This paper details the first results of a binary star observation project using the new high-resolution imaging camera FastCam mounted to the TCS at Teide Observatory. TCS is a 1.52-m infrared telescope installed at the Observatorio del Teide (OT) operated by the Instituto de Astrofísica de Canarias (IAC).

### 2.1 FastCam

FastCam (Oscoz et al. 2008; Rodríguez Ramos et al. 2008) is an instrument jointly developed by the spanish Instituto de Astrofísica de Canarias and the Universidad Politécnica de Cartagena. It was designed to obtain high spatial resolution images in the optical wavelength range from ground-based telescopes. The instrument consists of a very low noise and very fast readout speed EMCCD camera, capable of reaching the diffraction limit of medium-sized telescopes from 500 to 850 nm. The undisturbed images represent a small fraction of the observations. Therefore, a special software package has been developed to extract, from cubes of tens of thousands of images, those with better quality than a given threshold, following the "lucky imaging" technique (see, e.g., Tubbs et al. 2003; Law, Mackay & Baldwin 2006). The frame selection is done in parallel with the data acquisition at the telescope. The instrument makes use of an Andor iXon DU-897 back-illuminated system containing a 512x512 pixel frame transfer CCD detector from E2V Technologies. The pixel size is 16 microns and the DU-897 camera allows up to 30 exposures per second.

The theoretical diffraction limit at TCS has been reached in the I band (850 nm) of 0.15 arcsec and similar resolutions have also been obtained in the V and R bands.

For this survey we configured FastCam to take 30 millisecond frames (a few and very faint binaries were observed taking frames of 50, 70 and 100 ms). The averaged pixel scale was of 42.0 mas, giving a field of view of 21.5 x 21.5 arcsec.

**2.2 Reduc, SURFACE and wavelet process**

The position angles and angular distances were obtained with *Reduc*[2] software developed by Losse (2010). Reduc includes the SURFACE tool (Morlet & Salaman 2005; Morlet & Salaman 2006) which fits the parameters of a tridimensional mathematical surface which represents the map of illuminations produced by a binary star. This mathematical surface implies an empirical model of the illumination spreading produced by the passage of the light through the Earth atmosphere. The model function is the product of a Cauchy–Lorentz function and a Gauss–Laplace function. The last version of this software can be obtained from Florent Losse[3].

In order to accurately measure binaries with high $\Delta$mag we have used the wavelet transform of the *IRIS* program[4] (Buil 1991). This analysis decomposes the image into separate frames showing details of increasing scales. There are different approaches to do wavelet decomposition. *IRIS* utilizes the method based upon the well-known "à trous" approach (see Starck & Murtagh (1994)), which uses a dyadic wavelet to merge non_dyadic data in a simple and efficient procedure. This technique allows us to strongly enhance the presence of point-like sources in regions where the primary halo generally dominates.

# 3 OBSERVING LIST AND CALIBRATION

The observing list was composed of a majority of systems with angular separations between 0.2" and 1.5". We took into account that the TCS can point targets with zenital distances lower than 64º, with a small restriction towards the north and south where the telescope is limited due to the equatorial mount. Four binary star populations were observed: (1) neglected and not resolved orbital binaries during at least five years; (2) neglected physical binaries with important Keplerian motion; (3) unconfirmed Hipparcos binaries (HDS); (4) bright components of new binaries discovered by Benavides et al. (2010) for duplicity detection. The neglected physical binaries were included in the observational program if the orbital motion was greater than 30 degrees in position angle and/or a change of 30 percent in angular separation. It was surprising to find many orbital binaries with the last observation performed fifteen or more years ago. Several additional sets were added. These are close long neglected pairs with high differential photometry and binaries with orbits graded as 1 and 2. These systems were used to estimated errors or to calibrate the CCD images.

The pixel scale and image orientation were determined observing calibration binaries. These calibration binaries were chosen from the Sixth Catalog of Orbits of Visual Binary Stars (Hartkopf & Mason 2001) with orbits graded as 1 and 2 in this catalog. Note that utilizing double stars as calibrators for other double star measurements is far from optimal but TCS is not equipped with mask slits.

About 5000 individual frames were taken for each binary, although exceptionally, only 1000 frames were imaged for a very few bright targets. When the target is very weak, then 10,000 frames were taken. Each frame has an exposure time between 30 and 100 milliseconds. In order to construct a high resolution long exposure image, we aligned and averaged only those frames of better quality (about 10-25 per cent). So, the final image presents a total exposure time of 15-30 s.

From January 23, 2009 through November 27, 2009, the TCS was used on 10 scheduled nights (see Table 1) in 6 runs. Some nights were lost due to weather conditions and technical problems.

We obtained a total of 424 measurements for 157 stellar systems, which have been grouped into 198 mean positions. For many binaries, the digital images used in this work are the first ever taken (and therefore the measurements listed here are the first "digital" ones). Twenty binaries were not resolved due to large differential photometry and/or close separation. In addition, some components of new binaries discovered by Benavides et al. (2010) were observed.

# 4 METHOD OF ORBITAL CALCULATION

For several binaries, consistent residual trends demonstrated systematic runoff from published orbits. For some of these binaries, we were able to obtain new orbital elements that, although still preliminary, allow a better ephemeris estimate.

Before using any analytical method to calculate orbital parameters, θ was corrected for precession and θ and ρ were plotted against time which, allows the detection of measures with important errors or quadrant problems. Measures with the largest errors were assigned zero weight.

The orbital elements were calculated using Docobo's analytical method (Docobo 1985). It is briefly summarized in Docobo et al. (2000) and Tamazian et al. (2002). This method is applicable even when only a relatively short and linear arc of the orbit has been measured. The advantage of this method, over other formal solutions, is that it does not require knowledge of the areal constant. A family of Keplerian orbits is generated whose apparent orbits pass through three base points. Simultaneously, O-C (observed minus calculated) residuals in both ρ and θ are determined for these orbits.

The next step is to select the orbit with smallest weighted rms values for ρ and θ. Usually, the orbit with the smallest rms values in θ is not exactly the same as the orbit with the smallest rms values in ρ. In this case, we selected the orbit with minimum residuals calculated by the followed formula:

---

[2] http://www.astrosurf.com/hfosaf
[3] Email: florent_losse@yahoo.fr
[4] http://www.astrosurf.com/buil/us/iris/iris.htm

$$c^2 = \sum_{i=1}^{n} w_{qi}(r_i \Delta q_i)^2 + \sum_{i=1}^{n} w_{ri}\Delta r_i^2 \qquad (1)$$

where $w_{\theta i}$ and $w_{\rho i}$ are the weights for the i- $\theta$ and $\rho$ measures. $\Delta\theta$ (expressed in radians) and $\Delta\rho$ are the O-C residuals for $\theta$ and $\rho$.

The residual (O-C) for $\theta$ is expressed in radians. If Keplerian orbits show a flat rms gradient, then some orbits are rejected by means of the comparison of the dynamical mass, with that determined using spectral types.

The three base points have to be chosen carefully where the observational data seem most reliable with respect to instrumentation, data density, or critical arc coverage. We also tried to cover as much of the observed arc as possible. This may let the area around a single observation represent a base point without additional observational coverage.

The initial weights for measures were assigned using a data weighting scheme based on Hartkopf et al. (1989), Mason et al. (1999), Seymour et al. (2002) and Docobo & Ling (2003). The initial $\theta$ weights were five times larger than $\rho$ weights (Heintz 1978a) for visual measures.

After several iterations in the differential correction process, the measures with residuals larger than 3$\sigma$ were assigned zero weight. Later, the non-zero weight measures were reassigned following the work of Irwin et al. (1996).

## 5 RESULTS

Table 2 presents 424 mean relative positions of the components for 157 systems. The first two columns identify the system by providing the WDS designation (epoch 2000 coordinates) and discoverer designation. The total number of cubes measured is listed in column (3). We often recorded five data cubes for each object and they are analyzed independently. This strategy helped us to verify new companion detections and to estimate observational errors. Columns (4)–(6) give the epoch of the observation (expressed as a fractional Besselian year), the position angle (in degs), and the separation (in arcsecs). Note that the position angle has not been corrected for precession, and is thus based on the equinox for the epoch of observation. The errors reported in Columns (5) and (6) were computed by adding, quadratically, the calibration errors to the standard deviations of series of measurements obtained with the same data sets. Columns (7) and (8) give the WDS magnitudes for the primary and the secondary components. The last column shows the same notes. The most common note indicators are ''C'' indicating a confirming observation, and a number (N), indicating the number of years since the system was last measured. This is only given for systems with N>20 yr. Note that, in some cases, the time since the last observation is surprisingly large (for example, BU 1083 BC remained long neglected during 109 years or L 19, 111 years). Eighteen systems are confirmed here, and all of them are Hipparcos double stars. The angular separations measured range from 0.14" to 15.40", with a median separation of 0.52". The V magnitudes for the primary components range from 3.10 to 12.74 with a median of 9.00 magnitudes. The measured binary with the maximum photometric difference was HJ 2477 AB, with magnitudes of 3.1 and 10.1 respectively (about 7.0 magnitudes), separated 2.337". Figures 1 and 2 show, for the binaries measured in this work, the distribution of magnitudes for the primary components and the angular separations.

WDS 18040+3923 = L 19 is a binary discovered by Secchi (1860) in 1859.89 (321.1° and 0.45") composed of stars with magnitudes of 9.0 and 9.2 (WDS). It has 7 measurements, but the last one was performed in 1898.71, which is 111 years ago. For five measurements, the angular position ranges from 261.5 to 274.5 deg and the angular distance from 0.20 to 0.38 arcsecs. In this work, we imaged HIP 88488 (= HD 165434) with 7.5 magnitude in the V band and located at about 6.0 arc mins north from the WDS coordinate. We observed a very weak companion to HIP 88488 at 0.207 ± 0.012 arcsecs in the direction 273.9 ± 2.8 degrees. The CCD images were processed by a wavelet analysis due to the weakness of this companion. Our measurement is in good agreement with most of the historical ones. However the differential magnitude measured is $\Delta m = 2.5$. This value is not in agreement with the data listed in the WDS database. It is important to confirm the existence of this companion.

Table 3 lists unresolved systems, probably due to orbital or differential proper motion, making the separation between the components too small at the epoch of observation. In other cases a large $\Delta m$ caused the lack of splitting.

The first two columns in the same Table 3 identify the system by providing the WDS designation (epoch 2000 coordinates) and discoverer designation. Columns (3) and (4) give the epoch of our observation and the last WDS measurement (expressed as a fractional Besselian year). The number of observations listed in the WDS, the position angle (in degs) and the separation (in arcsecs) of the last WDS observation are listed in columns (5) to (7). Columns (8) and (9) give the WDS magnitudes for the primary and the secondary components. Some comments are shown in the last column.

### 5.1 New close companions

Benavides et al. (2010) reported the discovery of 142 new wide common proper motion binaries. We imaged same bright components of these wide binaries with the goal of detecting new close stellar companions. We imaged 14 components and for 11 of them, no stellar companion was detected. Table 4 lists these objects with no stellar companions. The first two columns list the AR and DEC (J2000); column (3) gives the discoverer designation and the component of the wide binary. Column (4) gives the HD or PPM identification; column (5) shows the epoch of our observation (expressed as a fractional Besselian year). Column (6) the V magnitude from WDS catalog.

Two new stellar companions were detected near to BVD 34 A and J 621 B. Table 5 lists photometry and position for these new companions. The first column gives the proposed name to these new objects; columns (2) and (3) give the magnitude in I band for the new companions and the instrumental differential magnitudes. Columns (4) and (5) show the position angle and distance to the nearest known stellar component; an approximated spectral type estimation is listed in column (6). This spectral type was estimated assuming that the new companion is at the same distance to the nearest component. The last column gives same notes.

The primary component of BVD 34 (= WDS03201+3611) has a stellar companion at 15.15" in the direction of 62.4°. This companion is the known B component of the BVD 34 system discovered by Benavides et al. (2010).

We need more astrometric measurements to determine if these new stellar companions are physically bound to the nearest stellar component, and we can only obtain information assuming a common distance. BVD 36 A (V = 9.93, $\mu(\alpha)$ = -0.9±1.0 mas yr$^{-1}$ and $\mu(\delta)$ = -39.3±1.0 mas yr$^{-1}$, 108.1 pc (Benavides et al. 2010)) is an G3V star with a magnitude of 9.24 in I band (we used a V-I = +0.70 for an G3V star). From the measured instrumental differential magnitude, the new companion is a star with I ~ 14.2 (we assumed that the observed instrumental magnitude difference is similar to $\Delta I$). If this new component is at the same distance that BVD 36 A, then it would be an M3V red dwarf. Astrometric measurements in the next 5-10 years is essential to determine the physical relation of this system. A relative motion with respect to BVD 36 A of about 0.2" in 5 years would imply that this system is a background star.

## 5.2 Radial velocities

Nowadays, radial velocities can be obtained from spectroscopic observations. This technique provides sufficient precision (a few km s$^{-1}$) to measure relative velocities in visual systems. In fact, some binaries in our sample show relative radial velocities between components large enough to be estimated spectroscopically. That is, these binaries could also be considered spectroscopic binaries. Radial velocities help us to improve the quality of preliminary orbital parameters and to remove the ascending node ambiguity. In this work, $\Delta$Vrad was calculated only for visual binaries with grade 4 and 5 orbits. $\Delta$Vrad ephemerides were calculated using the formulae (27) presented in Neckel (1986). Table 6 lists binaries with grade 4-5 orbits and $\Delta$Vrad > 5.0 km s$^{-1}$ with the hope that, in the future, other astrophysicists will obtain $\Delta$Vrad that help to constrain the orbital solutions for these visual binaries. The first two columns give the WDS designation and the discoverer designation. Columns (3) to (8) list the $\Delta$Vrad (in km s$^{-1}$), the epoch for the $\Delta$Vrad calculated, the V magnitudes for A and B components, the references for the orbital parameters and the grade of the orbit, respectively. Figure 4 shows a curve of relative radial velocity (in km s$^{-1}$) for HU 718 in 2010-2020. The orbital parameters published in Rica Romero (2010c) were used to calculate the radial velocities.

BU 1100 is classified by SIMBAD as a spectroscopic binary. In 2021, this stellar system will show a $\Delta$Vrad that ranges from -20 to -28 km s$^{-1}$. The binary system, A 750, will increase its $\Delta$Vrad up to +30.5 km s$^{-1}$ in 2013.34.

## 5.3 Measurement Capabilities

### 5.3.1 Internal errors

A good internal error estimate may be derived from the scatter in position angle and separation given in Table 2, columns (5) and (6). These errors were estimated, taking into account the calibration errors. The mean standard deviation in separation is found to be 16 ± 12 mas for binaries closer than 2" and 11 ± 6 mas for binaries closer than 0.5". The error in the tangential coordinate is 9 ± 4 mas. The internal error in position angle is 0.5 ± 0.2 °/$\rho$, where $\rho$ is the separation in arcsecs. If calibration errors are not considered, the radial and tangential errors have the same value (9 and 7 mas). However, if the calibration errors are taken into account, the tangential errors (i.e. errors in $\theta$ expressed in arcsecs) are smaller than radial errors (in $\rho$). This effect is not important for binaries with $\rho$ < 0.5", while for wider binaries, the effect is evident and is caused by the errors in calibration. For these wide binaries, the calibration errors are comparable to the total internal error in $\theta$ and $\rho$ because of the cumulative effect of the scale error. This fact causes errors in $\rho$ to be greater than the tangential error. For binaries with $\rho$ > 2" the internal errors are likely overestimated since the O-C (observed minus computed) residuals of our measures for the grade 2 orbital STF 2272 AB (with $\rho$ = 5.666") are +0.1° (= 0.011") and +0.019". Another example is the long period binary LDS 783, in which the orbit was computed in 2008. Our angular separation was of 2.503" with the residuals value of +0.1° (0.006") and -0.033". Table 7 summarizes the internal errors.

### 5.3.2 External errors

Binaries with well-characterized motion can be also be used to obtain an estimation of the external errors. Table 8 lists a summary of the external errors for our measures, the root mean square (rms) of residuals (O-C) and the averaged residuals for binaries with very well determined orbits (graded 1 and 2). To calculate these errors, 26 binaries were used. Before the external errors calculation, we eliminated the binaries whose orbits showed systematic runoff. The residuals in separation exhibit an rms of 23" mas. This means an error of 4%, while the tangential residuals have an rms of 0.013" (or 0.73 ° / $\rho$). The averaged residuals have an offset of nearly -0.01" in separation. The larger errors in the radial direction ($\rho$ residuals) and its negative offset may be due to proximity effects. The usual measure procedure in the REDUC software could pull the secondary peak toward the centre. This means a larger scatter in the radial direction, while the tangential component of the peak position is unaffected. This effect disappears with the SURFACE tool (within REDUC).

Figure 5 shows the O-C residuals for the observations of graded 1-2 orbits. The central rectangle shows $1\sigma$ bands in radial ($\rho$) and tangential ($\rho\Delta\theta$) directions at 0.023" and 0.013" according to the rms O-C residuals. The quality of the calibration orbit often produces large residuals. For some of these orbits, this could be the moment to update their elements. The binaries out of $3\sigma$ box are systems whith orbits showing systematic runoff with respect to recent measures or are suspected to show systematic runoff.

### 5.3.3 Limits of duplicity

Figure 6 plots the positive and negative observations in the $\rho-\Delta m$ space. The $\Delta m$ values came from the WDS catalog. WDS separations are also used for the negative observations. Table 9 lists a guideline with duplicity limits. Two negative observations must be detailed because they are in a region of the $\rho-\Delta m$ space where the binaries are resolved without problem. **WDS 00026-0829 = A 428:** It is a binary with components of 9.82 and 10.07 magnitudes ($\Delta V = 0.25$ mag.), which has not been resolved since 1991.25 (13° and 0.27"). The orbital ephemerides for the epoch of observation give 346.1° and 0.24", so we should be able to split it. On the other hand, **WDS 19153+2454 = HDS 2724** is a binary composed of stars of 10.36 and 10.82 magnitudes ($\Delta V = 0.46$ mag.). It has been measured three times, with the last time being in 2007.423 (295.8° and 0.16").

### 5.4 Orbit Calculations

Many of the binaries here listed have already calculated orbital parameters. Frequently, one measure may present important residual, so showing large differences between the observed and calculated positions. Eleven binaries have rho residuals greater than 0.20" and 17 binaries have theta residuals greater than 20°. In practice, these residuals may be caused by the low quality of the measures or orbital parameters estimations. However, to confirm the high quality of our measures we compute the residuals for all the historical measures for each binary. We obtain similar residuals in our measures respect recent and high quality historical ones. We also studied if our observations match with the evolution of the historical $\theta$ and $\rho$ values vs the time. In all cases, our measures present high quality. Therefore all these binaries with large residuals need orbital recalculations in a near future.

We present new orbital parameters and other results (residuals, ephemerides, masses, parallaxes, apparent orbits, etc.) obtained for four visual binary systems. These are WDS01032+2006 (LDS 873), WDS16492+4559 (BU 627 A-BC), WDS17184+3240 (BU 628), and WDS23114+3813 (HO 197 AB). Note that the BU 627 A-BC orbit is first published in this work. Initial orbits were calculated using the method of Docobo (1985) which were improved upon using the differential correction method of Heintz (1978a). Dynamical parallaxes were calculated using the Baize-Romani algorithm (Baize & Romaní 1946). The orbital parameters presented here have previously been published in the Information Circular of IAU Commission 26 (hereafter IAUDS).

Table 10 lists, for each binary, the date of its discovery, the final orbital elements, rms residuals and the mean absolute (MA) residuals. These are, in both cases, weighted averages calculated using the data-weighting scheme described in section 4. The $a^3/P^2$ values are also listed. Errors for the elements were computed making use the covariance matrix and the residuals to all observations.

Table 11 presents ephemerides for the period 2011-2020. Table 12 lists the stellar data and is organized as follows: WDS magnitudes and spectral types in columns (3)-(5); in column (6), the dynamical parallaxes calculated using the orbital periods and semimajor axis obtained in this work; in columns (7) and (8) the apparent magnitudes published in the Hipparcos and Tycho catalogs (ESA 1997), in column (9) the corresponding trigonometric parallaxes (with their standard errors); and in column (10) the total mass of the binary (± estimated standard error), as calculated from the Hipparcos parallax.

Figures 7-10 show the new apparent orbits drawn together with the observational data.

## 6 DESCRIPTION OF SOME SYSTEMS

### 6.1 Orbital Binaries

In this section, photometric, astrometric, spectroscopic and kinematical data are analysed. Spectral types and luminosity classes, masses and dynamical parallaxes are also obtained. The dynamical parallax was calculated using the Baize-Romaní method (Baize & Romaní 1946). We also made use of the grading scheme from the Sixth Catalog of Orbits of Visual Binary Stars (Hartkopf, Mason, & Worley 2001).

The absolute magnitudes were estimated using the tables published by Zombeck (1990), which relate spectral types with absolute magnitudes. For red stars (V-K = 3.33) we related optical-infrared colors with absolute magnitudes derived from Henry et al. (1997). We assumed that most of these stars are near solar metallicity.

A kinematical study was realized to determine the stellar age. From galactocentric velocity (U, V, W), we use the Eggen (1969a, 1969b) and Chiba & Beers (2000) diagrams in addition to the kinematic age parameter of Grenon (1987), fG. Bartkevicius & Gudas (2002) determined the relation between fG and the age. Statistically, the stars with fG < 0.20 belong to the young-middle age group (with an age less than 3-4 Gyr) of the thin disk population. The stars with 0.20 < fG <

0.35 belong to the old (with age of 3-10 Gyrs) thin disk population. The stars with 0.35 < fG < 0.70 belong to the thick disk population (age greater than 10 Gyrs) and the stars with fG > 0.70 belong to the halo population.

In order to check the membership of the binaries to young kinematic groups, we consulted Table 1 published in Montes et al. (2001) and Soderblom & Mayor (1993).

Table 13 lists the main astrophysical features for the binary system.

**WDS 01032+2006 = LDS 873:** Since Luyten (1969) discovered it in 1930, LDS 873 (= GJ 1026 = NLTT 406-76/77 = LHS 1180) accumulated 18 measures, which cover an arc of 25 degrees. Recently, Lampens & Strigachev (2001) performed 4 observations in 1999 using the 1.3-m telescope at Skinakas Observatory. The orbital parameters presented in this work were published in Rica Romero (2008).

LDS 873 is composed of stars with 11.68 and 12.97 Hipparcos magnitudes and M1.5V and M3.5V spectral types. The secondary component is listed in "Catalogue and Bibliograhy of UV Cet stars" (Gershberg et al. 1999).

From BVIJHK photometry, we estiamte M1.5V and M3V spectral types, which is in very good agreement with the literature.

From the relation between spectral types and masses (Kirkpatrick & McCarthy (1994)) we can conclude that LDS 873 is a system with two 0.45 and 0.26 $M_\odot$ stars. The sum of the masses is in agreement with the dynamic mass within errors.

Seymour et al. (2002) calculated the previous orbit. There have been no new measures since then, but the measures did not obey the law of areas and the mass of the system calculated from dynamical parameters, which was about 20 $M_\odot$, larger than expected for two M red dwarfs. A large series of measures made by Heintz (1978b) showed errors in the WDS database. The decimal points were offset one position to the right (e.g. the theta value 24.78 measured by Heintz in 1940.74 appeared in WDS database as 247.8 degrees). This error has been corrected in WDS database.

**WDS 16492+4559 = BU 627 A-BC:** It is composed of stars with 4.87 and 8.85 Hipparcos magnitudes and spectral types A2V and G7/8V. Since Burnham (1879) discovered it in 1878, BU 627 A-BC (= 52 Her = V 637 Her = ADS 10227) accumulated 54 catalogued measurements which cover an arc of about 92 degrees. The last one was made in 1997.54 by Courtot (1998) with a result of 35.9º and 2.19". In this work, we present a new measurement (see Table 2). This is the first orbit published for this binary. The orbital parameters presented in this work were published in Rica Romero (2010b).

From the absolute magnitude, a total mass of 4.03 $M_\odot$ was estimated in agreement with the dynamic mass.

The (U, V, W) galactic heliocentric velocity is very similar to those stars members of the UMa group (Sirius Supercluster). In the literature, this binary is considered as a member of the UMa Group (Geary & Abt (1970), Levato & Abt (1978), Hubrig & Schwan (1991)). Therefore this system has an age of about 300 Myr. We searched for new proper motion companions. No new companions were found.

However, one question still remains: Is BU 627 A-BC a bound stellar system? The trajectory of the secondary component shows a linear path. Therefore, it is important to confirm the physical nature of this stellar system before the orbital calculations. In this case, we have performed several tests. These tests were detailed described in Benavides et al. (2010).

Using historical measures, a weighted linear fit was performed to calculate a relative motion of the BC components with respect to A. We obtained $\Delta x = +21.2 \pm 0.2$ mas yr$^{-1}$ and $\Delta y = +4.4 \pm 0.2$ mas yr$^{-1}$. The time baseline of the 55 historical measures was 131 years.

According to the test published by **Dommanget** (1955, 1956), the stellar system will be physically bound when it is located nearer than 72 pc. The system is located at 55 pc therefore, from this point of view, this means the system is physically bound. In the **van de Kamp test** (1961), the condition for a parabolic orbit is the critical value to determine if the relative velocity of the system is periodic or non-periodic. The true critical value for a parabolic orbit is 340.3 AU$^3$ yr$^{-2}$ while the observed projected critical value is 197.8 AU$^3$ yr$^{-2}$. Therefore, BU 627 A-BC may be a physical pair. The test presented in **Sinachopoulos & Mouzourakis test** (1992) studies the compatibility of the observed relative proper motion with that dynamically allowed. For this system, the tangential velocity of the secondary star, with respect to the primary, is 5.64 km s$^{-1}$. On the other hand, the maximum orbital velocity using Kepler's Third Law is 5.87 km s$^{-1}$. In short, the three tests performed seem to point to the same fact, i.e. that BC is gravitationally bound to BU 627 A.

**WDS 17184+3240 = BU 628:** It is composed of two stars with 9.26 and 10.20 magnitudes (ESA 1997), in the Hipparcos photometric band, and a combined spectral type of G0. Since Burnham (1879) discovered this system in 1878, BU 628 (= ADS 10459) has had 64 catalogued measurements which cover an arc of about 110 degrees. The last observation was made in 1997.61 by Heintz (1998). Here, we present a new measurement (see Table 2) and orbital parameters.

An orbit was calculated in the past by Zulevic (1986), but speckle measures showed important residuals for this orbit. In comparison with the orbit of Zulevic, our orbit has a larger orbital period and semimajor axis. We obtained 699 years and 0.885" (see Rica Romero 2009), for these metioned parameters, respectively. We searched for new proper motion companions, but no new companions were found separated more than 0.15 arcsecs.

**WDS 23114+3813 = HO 197 AB:** It is composed by two stars of 8.47 and 9.16 magnitudes (ESA 1997) in the Hipparcos photometric band. Since Hough (1887) discovered this system in 1885, HO 197 AB (= ADS 16576 = HD 218917) accumulated 59 catalogued measurements which cover an arc of about 190 degrees. The last available measurement was obtained in 2003.717 by Salaman et al. (2005), with result of 308.0º and 0.30". In this work, we present two new measurements (see Table 2). Since 2005 the relative positions have significantly changed.

The previous orbit was calculated by Docobo & Costa (1990), but our measures show important residuals (about 20-22 degrees in theta and about -0.25 arcsec in rho) for this orbit. We present new orbital parameters for HO 197 AB listed in Table 10. These orbital parameters were published in Rica Romero (2010a). We also have obtained a combined spectral type of F5V using BVIJHK photometry. Our spectral type is in agreement with the literature. Differential photometry from Hipparcos satellite is +0.69. Using the combined spectral type and $DV$ (calculated from $\Delta Hp$ using Table 1 presented by Bessell 2000), we obtained individual spectral types of F4V and F7V (assuming a dwarf nature). From Hipparcos data, we computed absolute magnitudes of +2.25 and +2.94 (far brighter than the typical absolute magnitudes for F4V and F7V). Using isochronas generated by CMD 2.3 (http://stev.oapd.inaf.it/cmd) we determined that both components are surely evolved stars.

Our dynamic total mass is very preliminary ($12.0 \pm 7.0$ $M_\odot$) and it is much higher than expected for two evolved stars with about 2 Gyr (about 3.9 total mass). We think that the preliminary orbital parameters is the cause for this disagreement. Regarding new proper motion companions, no new ones were found more separated than the resolution limit of FastCam.

**6.2 Wide Components of WDS 16492+4559**

The close pair BU 627 AB has three wide components listed in the WDS catalog. Components D (at 67"; 12.4 magnitude) and E (143"; 12.6 magnitude) were added to the system WDS 16492+4559 in 1881.42 by Bigourdan (1883) when he performed a micrometric measurement using a 12 inch telescope. The more recent measurement came from the 2MASS project in 1998.43, with results of 235.1º and 65.26" for AD and 269.6º and 146.95", for AE).

The F component was added to the system in 1984.318 by Kazeza (1984) when he performed a photographic measurement using a 13 inch telescope. This component is listed in the WDS catalog as a star of 13.9 photographic magnitude. The more recent measurement came from the 2MASS project in 1998.43 (252.5º and 91.49").

The WDS catalog lists visual magnitudes of 11.8, 12.7 and 13.9 for the D, E and F components, respectively. In order to obtain a more accurate estimate for the V magnitude for these wide components, photometric data of the CMC14 (CMC 2006) and UCAC3 catalogs (Zacharias et al. 2009) were used. The process to derive the V magnitude was published in Pavlov (2009) and Rica Romero (2010d, section 5.4).

The final V values were $12.39 \pm 0.08$, $12.61 \pm 0.06$ and $14.13 \pm 0.08$ for the D, E and F components, respectively. These magnitudes were determined calculating a weighed average for the V values from UCAC3 and CMC14 data. The values for E and F are in good agreement with those of the WDS catalog.

No spectral type information was found in the literature for these components. From V, J, H and K magnitudes, we determined that the D component is likely a K2V star located at about $161 \pm 32$ pc. The giant nature was rejected because, then, this star would have an improbable tangential velocity ($> 200$ km s$^{-1}$). The E component seems to be a K8III giant star located very far away, at about $3800 \pm 760$ pc. The F component is an F7 star. For the distance values, a 20% error was assumed (see Benavides et al. 2010). The galactic latitude for this stellar system is +40 degrees. The determined spectral types were not corrected by the reddening calculated in this work (E(B-V) <= 0.02).

The distances of D and E are incompatible with that for AB and so, for sure, they are not bound to AB.

In the literature there is no kinematical information for the D, E and F components. Using the historical measures, a weighted linear fit was performed to estimate the relative motion of the wide components with respect to AB:

For D: $\Delta x = -36.2 \pm 1.4$ mas yr$^{-1}$ and $\Delta y = +66.7 \pm 2.0$ mas yr$^{-1}$
For E: $\Delta x = -22.4 \pm 5.0$ mas yr$^{-1}$ and $\Delta y = +52.3 \pm 10.0$ mas yr$^{-1}$

From the proper motion of AB and the relative motion for the wide components, we calculated their proper motions (see Table 14). For the F component, the WDS catalog lists only two measurements, performed in epochs 1984.318 and 1998.43. There are not enough measurements and the time baseline is very small. So, we cannot obtain an accurate relative proper motion. Therefore, we did not determine it in this work. Due to the lack of accurate and good enough data, we could not determine if the F component is bound to the AB close pair or not.

Proper motions for D and E are very different and incompatible (much greater than 3 σ) with the proper motion of AB. Therefore components D and E are surely not gravitationally bound to AB i.e. they are optical companions. These pairs must be flagged with the "U" code ("*Proper motion or other technique indicates that this pair is non-physical.*") in the Note column of the WDS Index Catalog.

**7 CONCLUSIONS AND FUTURE WORKS**

This work is the first part of a long-term program of binary-systems carried out with the TCS telescope and the lucky-imaging FastCam camera. We have presented 424 lucky-imaging observations of 157 binary stars taken in 2009 with the FastCam camera mounted at the 1.5-m TCS telescope of the Teide Observatory (Spain). Some of these objects have been resolved for the first time using CCD techniques (speckle, lucking-imaging, adaptative optics, etc). We confirmed 18 Hipparcos binaries and discovered two new companions to J 621 B and BVD 36 A. The measures presented in this work will help to determine or refine orbits for these objects in the future. In fact, we recalculated orbital parameters for three binaries (LDS 873, BU 628 and HO 197 AB) and we presented, for the first time, the orbital parameters for BU 627 A-BC.

The calculated internal errors (calculated as the spread of several measures) were 7 mas in ρ and 1.2º (9 mas) in θ. The external errors were calculated comparing our measures with the ephemerides of well-known orbital binaries. The rms deviation in ρ was 23 mas and in θ 0.73º/ρ. A study of the duplicity limits yields values of about a maximum ΔV ~ 6.0 for binaries wider than 2", and a ΔV~2.0 for binaries with angular separation of about 0.2".

## ACKNOWLEDGMENTS

This article is based on observations made with the 1.52-m Carlos Sánchez Telescope operated on the island of Tenerife by the Instituto de Astrofisica de Canarias in the spanish Observatorio del Teide. This research has made use of the Washington Double Star Catalog maintained at the U.S. Naval Observatory.

We thanks to the referee and editor for comments on this article. Their suggestions have contributed significantly to the improvement of this work.

We kindly acknowledge Dave Arnold for the English Grammar revision of this work and Edgardo Masa for a general review.

**Table 1**. Observational runs

| Runs during 2009 | Num. Binaries / Calibrations | Pixel scale [mas pixel$^{-1}$] | Position Angle offset [deg] |
|---|---|---|---|
| 23/01 [1] | 18 / 3 | 41.77 ± 0.33 | 1.35 |
| 24/03 [2] | 17 / 5 | 42.17 ± 0.77 | -0.39 ±1.02 |
| 09/06 [2] | 22 / 3 | 41.91 ± 0.42 | 1.07 ± 0.75 |
| 13-14/08 | 24 / 4 | 41.89 ± 0.38 | -0.55 ± 0.61 |
| 25, 27/09 & 02/10 | 80 / 3 | 42.13 ± 0.57 | -0.38 ± 0.68 |
| 27, 28/11 | 75 / 7 | 42.06 ± 0.86 | -0.69 ± 0.94 |

(1) Some hours were lost by high humidity.
(2) Some hours were lost for testing FastCam.

**Table 2.** Lucky Imaging Measurements of Double Stars

| WDS designation α, δ (2000) | Discoverer Designation | Num. Cubes | Epoch (2000+) | θ (°) | ρ (") | mg. A WDS | mg. B WDS | Notes |
|---|---|---|---|---|---|---|---|---|
| 00022+2705 | BU   733 AB | 1 | 9.7550 | 269.17 | 0.821 | 5.83 | 8.9 | |
| 00022+2705 | BU   733 AB | 1 | 9.9081 | 273.00 | 0.820 | 5.83 | 8.9 | |
| 00078+6321 | HU 1002 | 1 | 9.7550 | 275.76 | 0.429 | 9.22 | 9.5 | |
| 00283+6344 | HU 1007 | 1 | 9.7550 | 91.46 | 0.408 | 10.25 | 10.13 | |
| 00303+5959 | BU 1094 AB | 1 | 9.7550 | 290.33 | 0.258 | 6.07 | 8.45 | |
| 00364+1213 | A    807 | 1 | 9.7550 | 230.67 | 0.789 | 9.58 | 10.07 | |
| 00429+5742 | A    916 | 1 | 9.7550 | 240.00 | 0.277 | 10.11 | 10.28 | |
| 01014+3535 | COU 854 | 1 | 9.9109 | 214.64 | 0.154 | 9.95 | 9.52 | |
| 01032+2006 | LDS 783 | 1 | 9.7550 | 49.06 | 2.439 | 12.24 | 12.97 | |
| 01032+2006 | LDS 783 | 1 | 9.9082 | 53.81 | 2.503 | 12.24 | 12.97 | h |
| 01148+6056 | BU 1100 | 1 | 9.7550 | 356.02 | 0.304 | 8.15 | 8.07 | a |
| 01148+6056 | BU 1100 | 1 | 9.9082 | 356.18 | 0.257 | 8.15 | 8.07 | |
| 01157+5918 | A    935 | 1 | 9.7550 | 344.21 | 0.215 | 8.96 | 9.89 | |
| 01157+5918 | A    935 | 1 | 9.9082 | 344.52 | 0.217 | 8.96 | 9.89 | |
| 01158+0947 | A   2102 | 1 | 9.7550 | 114.94 | 0.314 | 7.36 | 9.89 | |

NOTES. -- *Table 2 is published in its entirety in the electronic edition of the MNRAS. A portion is shown here for guidance regarding its form and content.* (a) magnitudes from Hipparcos catalog; (h) Orbital parameters calculated in this work. Residuals: +0.1° and -0.03".

**Table 3.** Binaries not resolved

| WDS designation α, δ (J2000)) | Discoverer Designation | Epoch (2000+) | Epoch WDS | N | θ (°) | ρ (") | mg. A WDS | mg. B WDS | Notes |
|---|---|---|---|---|---|---|---|---|---|
| 07322+1405 | HU 1244 | 9.063 | 1992.14 | 39 | 281.5 | 0.28 | 10.45 | 10.36 | a |
| 20496+5047 | HDS2968 | 9.621 | 1991.25 | 1 | 106.0 | 0.30 | 7.31 | 10.67 | b |
| 02512+0141 | A 2338 | 9.736 | 2008.77 | 11 | 328.6 | 0.10 | 9.90 | 10.60 | c |
| 03272+0944 | HDS 433 | 9.737 | 2008.77 | 12 | 126.6 | 0.19 | 3.74 | 7.55 | b, d |
| 19153+2454 | HDS2724 | 9.736 | 2007.42 | 3 | 295.8 | 0.16 | 10.36 | 10.82 | b |
| 01418+4237 | MCY 2 | 9.736 | 2007.80 | 6 | ... | 0.05 | 4.95 | 11.00 | c |
| 22052+2952 | A 1451 | 9.741 | 1985.48 | 5 | 6.8 | 0.32 | 7.90 | 7.90 | e |
| 20496+5047 | HDS2968 | 9.741 | 1991.25 | 1 | 106.0 | 0.30 | 7.31 | 10.67 | b |
| 21296+4626 | HDS3059 | 9.741 | 1991.25 | 1 | 217.0 | 0.40 | 6.80 | 10.84 | b |
| 19296-1239 | HU 75 | 9.741 | 2009.27 | 63 | 60.8 | 0.16 | 7.09 | 9.15 | b |
| 23587-0333 | BU 730 | 9.755 | 2008.89 | 26 | 322.6 | 0.78 | 4.90 | 8.90 | f |
| 00258+1025 | HDS 57 | 9.755 | 2007.82 | 3 | 99.0 | 0.12 | 9.71 | 10.71 | c |
| 00026-0829 | A 428 | 9.908 | 1991.25 | 39 | 13.0 | 0.27 | 9.82 | 10.07 | f, g |
| 04362+0814 | A 1840 AB | 9.909 | 2008.77 | 40 | 17.3 | 0.15 | 8.37 | 8.91 | b |
| 02512+0141 | A 2338 | 9.911 | 2008.77 | 11 | 328.6 | 0.10 | 9.90 | 10.60 | c |
| 23304+3050 | BU 1266 AB | 9.911 | 2007.60 | 149 | 260.6 | 0.11 | 8.35 | 8.14 | c |
| 06120+0723 | HDS 846 | 9.909 | 1991.25 | 1 | 268.0 | 0.34 | 7.07 | 10.42 | b |
| 06201-0752 | HDS 866 | 9.909 | 1991.25 | 1 | 98.0 | 0.16 | 7.36 | 8.42 | b |
| 06500+4611 | HDS 948 Aa-Ab | 9.912 | 2007.80 | 2 | 125.8 | 0.35 | 7.38 | 10.2 | b |
| 22409-0333 | KUI 114 | 9.908 | 2006.56 | 82 | ... | 0.05 | 6.52 | 8.63 | c |
| 17348+0601 | HDS2481 Aa-Ab | 9.440 | 2006.56 | 2 | 246.7 | 0.10 | 7.76 | 9.23 | c |

**NOTES. --** **(a)** In 2010.0646 (275.3° and 0.326"; unpublished measurement), we could resolve it with the same instrumental setup used in this work; **(b)** At the limit of the instrumentation; **(c)** Angular separation smaller than the diffraction limit; **(d)** At the moment of observation the ephemerides for the orbit of Cvetkovic (2010) is 151.8° and 0.20"; **(e)** Not resolved by Couteau (1988) and Alzner (2003, 2005, 2008); **(f)** It should be resolved with the instrument used; **(g)** Ephemerides from the orbit of Zulevic (1984) give 346.1° and 0.239" (2009.9082).

**Table 4.** Observed Single Stars

| AR 2000 | DEC 2000 | WDS Designation | Designation 2 (PPM, HD) | Epoch (2000+) | V (mag) |
|---|---|---|---|---|---|
| 03 11 24.75 | +44 29 56.6 | POP 223 A | ... | 9.736 | 9.79 |
| 01 20 21.31 | +09 36 43.7 | BVD 15 A | PPM 144495 | 9.755 | 9.62 |
| 01 56 47.20 | +23 03 03.9 | BVD 21 A | HD 11850 | 9.755 | 7.86 |
| 23 10 29.36 | +41 19 19.5 | BVD 142 C | HD 218805 | 9.755 | 7.75 |
| 01 35 01.68 | +60 46 45.1 | HJL 20 A | PPM 13089 | 9.755 | 8.56 |
| 01 35 07.62 | +60 46 53.6 | HJL 20 B | PPM 13091 | 9.755 | 9.08 |
| 03 17 42.93 | +58 46 57.9 | BVD 32 A | HD 20208 | 9.908 | 8.14 |
| 05 09 12.57 | +11 29 43.2 | BVD 50 A | HD 33221 | 9.910 | 8.31 |
| 05 09 14.61 | +11 29 35.8 | BVD 50 B | HD 33234 | 9.911 | 8.46 |
| 06 52 27.42 | +59 39 07.9 | BVD 61 A | HD 49496 | 9.912 | 10.32 |
| 23 10 29.36 | +41 19 19.5 | BVD 142 A | HD 218805 | 9.911 | 7.75 |

**Table 5.** New Companions Detected

| Name | I mag. | Δmag | θ (°) | ρ (") | SpT | Notes |
|---|---|---|---|---|---|---|
| FMR 20 Aa,Ab | 14.2 | +5.0 | 243.8 | 0.97 | M3V | 1 |
| FMR 21 BD | --- | +2.2 | 24.5 | 1.16 | --- | 2 |

**NOTES.—** (1) New companion to BVD 36 A; (2) New companion to J 621 B. J 621 (=WDS 22547+1812) is a triple system composed of a close pair, AB, with magnitudes of 12.61 and 12.74 (WDS), separated by about 2.5" (see Figure 3) and a wider component, C, of 11.0 magnitude at 15.315".

**Table 6.** Relative Radial Velocities

| WDS designation α, δ (J2000) | Discoverer Designation | ΔVrad (km s$^{-1}$) | Epoch | Mg. A | Mg. B | References | Grade |
|---|---|---|---|---|---|---|---|
| 00283+6344 | HU 1007 | +5.0 | 2012 | 10.25 | 10.13 | Dsl1968 | 5 |
| 01014+3535 | COU 854 | -5.3 | 2012 | 9.95 | 9.52 | Hrt2009 | 3 |
| 01148+6056 | BU 1100 | -7.5 | 2012 | 8.15 | 8.07 | Mlr1955b | 4 |
| 09005+3225 | HU 718 | +19.9 | 2012 | 9.3 | 9.49 | Hei1997 | 4 |
| 18146+0011 | STF2294 | +7.7 | 2012 | 8.22 | 8.55 | Luy1934a | 4 |
| 20198+4522 | STT 406 | -24.0 | 2028 | 7.25 | 8.74 | Hei1976 | 3 |
| 20527+4607 | A 750 | -9.5 | 2012 | 9.11 | 10.26 | Hei1986b | 3 |
| 22375+2356 | HU 391 AB | -10.4 | 2011 | 10.14 | 10.89 | Hrt2009 | 4 |
| 22385+0218 | HO 479 | -8.2 | 2031-2034 | 8.50 | 9.75 | USN2007b | 3 |
| 23026+4245 | BU 1147 AB | +10.8 | 2018 | 5.19 | 7.70 | FMR2008b | 3 |
| 23114+3813 | HO 197 AB | -8.7 | 2012 | 8.47 | 9.16 | FMR2010 | 3 |
| 23409+2022 | HO 303 AB | +15.0 | 2060 | 8.50 | 10.81 | Hei1995 | 4 |

**NOTES.-- Dsl1968** (da Silva & Balca 1968), **Hrt2009** (Hartkopf & Mason 2009), **Mlr1955b** (Muller 1955), **Hei1997** (Heintz 1997), **Luy1934a** (Luyten 1934), **Hei1976** (Heintz 1976), **Hei1986b** (Heintz 1986), **FMR2008b** (Rica Romero 2008), **FMR2010** (Rica Romero 2010a), **Hei1995** (Heintz 1986), **USN2007b** (Brendley & Mason 2007).

**Table 7.** Summary of internal errors

| | $\sigma_\theta$ (degs) | $\rho\,\sigma_\theta$ (mas) | $\sigma_\rho$ (mas) |
|---|---|---|---|
| *Considering calibration errors* | | | |
| ρ < 0.5" | 1.5±1.0 | 8±4 | 11±6 |
| 0.5 <= ρ <= 2.0" | 1.0±0.3 | 12±4 | 25±16 |
| ρ > 2.0" | 1.0±0.1 | 142±77 | 383±209 |
| *Discarding calibration errors* | | | |
| Within the same night | 0.9±1.0 | 6±4 | 6±4 |
| In different nights | 1.2±0.9 | 9±8 | 7±6 |

**Table 8.** Summary of external errors

| **rms:** | |
|---|---|
| $\theta_{O-C}$ | 1.87° |
| $\rho \cdot \theta_{O-C}$ | 0.013" |
| $\rho_{O-C}$ | 0.023" |
| **Averaged residuals:** | |
| $\theta_{O-C}$ | +0.34±1.88° |
| $\rho\,\theta_{O-C}$ | +0.002±0.013" |
| $\rho_{O-C}$ | -0.009±0.022" |

**Table 9**. Duplicity Limits

| For binaries with… | Maximum observed $\Delta V$ |
|---|---|
| $\rho \sim 0.17"$ | ~0.75 mag. |
| $\rho \sim 0.20"$ | ~2.0 mag. |
| $\rho \sim 0.25"$ | ~2.5 mag. |
| $0.5" = \rho < 1.0"$ | ~4.0 mag. |
| $1.0" = \rho < 2.0"$ | ~5.0 mag. |
| $\rho = 2.0"$ | ~6.0 mag. |

**Table 10.** Orbital parameters, parallaxes and residuals

| Name | BU 627 A-BC | BU 628 | HO 197 AB | LDS 873 |
|---|---|---|---|---|
| WDS | 16492+4559 | 17184+3240 | 23114+3813 | 01032+2006 |
| ADS | 10227 | 10459 | 16576 | … |
| HIP | 82321 | 84653 | 114504 | 4927 |
| Disc. date | 1878.38 | 1878.41 | 1885.81 | 1936 |
| P (years) | 1977 ± 209 | 699 ± 11 | 170.3 ± 1.6 | 521 ± 9 |
| T (BY) | 1966.0 ± 5.2 | 1807.2 ± 3.3 | 1843.59 ± 1.70 | 2079.3 ± 0.5 |
| E | 0.329 ± 0.054 | 0.371 ± 0.010 | 0.456 ± 0.044 | 0.711 ± 0.008 |
| a (arcsen) | 4.914 ± 0.396 | 0.885 ± 0.050 | 0.419 ± 0.033 | 3.366 ± 0.012 |
| i (degs) | 63.8 ± 3.3 | 116.8 ± 1.6 | 106.7 ± 3.9 | 53.2 ± 1.0 |
| $\omega$ (degs) | 280.2 ± 2.4 | 325.8 ± 1.6 | 249.6 ± 3.6 | 81.0 ± 0.9 |
| $\Omega$ (degs) | 77.6 ± 5.2 | 27.3 ± 2.6 | 120.4 ± 3.6 | 92.5 ± 0.9 |
| Residuals: | | | | |
| rms($\theta$)(degs) | 2.66 | 1.62 | 2.74 | 0.43 |
| rms($\rho$) (arcsec) | 0.097 | 0.020 | 0.015 | 0.031 |
| MA($\theta$)(degs) | 2.04 | 1.04 | 1.83 | 0.38 |
| MA($\rho$) (degs) | 0.057 | 0.009 | 0.009 | 0.023 |
| $a^3/P^2$ (arcsec$^3$ yr$^{-2}$) | 3.03511·10$^{-5}$ | 1.60195·10$^{-6}$ | 6.05602·10$^{-6}$ | 4.16776·10$^{-5}$ |

**Table 11.** Ephemerides

| | 01032+2006 | | 16492+4559 | | 17184+3240 | | 23114+3813 | |
|---|---|---|---|---|---|---|---|---|
| Epoch | $\theta$ (deg) | $\rho$ (arcsec) | $\theta$ (deg) | $\rho$ (arcsec) | $\theta$ (deg) | $\rho$ (arcsec) | $\theta$ (deg) | $\rho$ (arcsec) |
| 2011.0 | 54.2 | 2.524 | 36.9 | 2.009 | 266.0 | 0.540 | 280.5 | 0.153 |
| 2012.0 | 54.7 | 2.513 | 37.4 | 2.021 | 265.4 | 0.543 | 275.8 | 0.134 |
| 2013.0 | 55.3 | 2.502 | 37.8 | 2.034 | 264.9 | 0.547 | 269.7 | 0.116 |
| 2014.0 | 55.8 | 2.490 | 38.3 | 2.047 | 264.3 | 0.551 | 261.3 | 0.098 |
| 2015.0 | 56.3 | 2.478 | 38.7 | 2.060 | 263.7 | 0.554 | 249.5 | 0.082 |
| 2016.0 | 56.9 | 2.466 | 39.1 | 2.073 | 263.2 | 0.558 | 233.0 | 0.071 |
| 2017.0 | 57.4 | 2.454 | 39.6 | 2.086 | 262.7 | 0.561 | 212.6 | 0.067 |
| 2018.0 | 58.0 | 2.442 | 40.0 | 2.099 | 262.1 | 0.565 | 192.1 | 0.071 |
| 2019.0 | 58.5 | 2.429 | 40.4 | 2.113 | 261.6 | 0.569 | 175.5 | 0.082 |
| 2020.0 | 59.1 | 2.416 | 40.8 | 2.126 | 261.1 | 0.572 | 163.6 | 0.098 |

**Table 12.** Stellar Data

| Star | | | WDS | | Present Work | Hipparcos | | | |
|---|---|---|---|---|---|---|---|---|---|
| WDS | HIP | $m_A$ | $m_B$ | Spectral Type | $\pi$ (Dynamical) (arcsec) | HpA | HpB | $\pi$ (Trigonometric) (mas) | $\Sigma$ (M$_\odot$) |
| 01032+2006 | 4927 | 12.24 | 12.97 | … | 65.6 | 11.68 | 12.97 | 61.23 ± 5.26 | 0.61 ± 0.16 |
| 16492+4559 | 82321 | 4.84 | 8.45 | A1V | … | 4.87 | 8.85 | 18.10 ± 0.34 | 5.1 ± 1.7 |
| 17184+3240 | 84653 | 9.48 | 9.56 | G0 | … | 9.26 | 10.20 | 7.10 ± 1.36 | 4.1 ± 2.5 |
| 23114+3813 | 114504 | 8.47 | 9.16 | F5 | … | 8.47 | 9.16 | 5.96 ± 1.06 | 12.0 ± 7.0 |

**Table 13**. Astrophysical Data for Orbital Binaries

|  | WDS 1032+2006 LDS 873 | WDS 6492+4559 BU 627 A-BC | WDS 17184+3240 BU 628 | WDS 23114+3813 HO 197 AB |
|---|---|---|---|---|
| Hp mag. | 11.62, 13.02 [a] | 4.87 , 8.85 | 9.26, 10.20 | 8.47, 9.16 |
| $(\mu_\alpha, \mu_\delta)$ [mas yr$^{-1}$] [b] | (+670.0 , +41.5) | (+22.1, -56.1) | (-25.1, -42.7) | (+51.6, -2.3) |
| Mv [c] | +10.57 , +11.97 | +1.13, +5.44, +5.54 | +3.52, +4.46 | +2.25, +2.94 |
| Distance [pc] [c] | $16.2^{+0.8}_{-0.9}$ | $55.2^{+1.1}_{-1.0}$ | $140.8^{+33.4}_{-22.6}$ | $167.8^{+36.3}_{-25.3}$ |
| Spectral Type | M1.5 – M3.5 [d] M2 [e] M2.0 [f] | A1V [h] A3V [i] A5 [j] | G0 [k] F7V, G3V [z] | F4V [m] F5 [n] |
| Stellar Mass [$M_{sun}$] | 0.45, 0.26 [z] | 2.32, 0.86, 0.85 [z] | … | … |
| Radial Velocity [km s$^{-1}$] | -3.8 [g] | … | … | … |
| (U,V,W) Velocity [km s$^{-1}$] | (-42, +29, +7) [z] | (+11, -4, +10) [z] | … | (-16, -42, +1) [l] |
| Stellar Population | Young galactic disk | Young galactic disk | … | Old galactic disc |
| fG | 0.22 | 0.08 | … | 0.23 |
| Age (Gyr) | … | … | … | $2.2^{+0.2}_{-0.2}$ [l] |
| [Fe/H] | … | … | … | -0.11 [l] |

NOTES.- **a)** Lampens & Strigachev (2001); **b)** Tycho-2 catalog (Høg et al. 2000); **c)** calculated using Hipparcos data; **d)** Gliese & Jahress (1991); **e)** Spectral type for the primary component (Reid, Hawley, & Gizis 1995); **f)** combined spectral type (Reid 2004); **g)** Reid et al. (1995); **h)** Abt (1981); **i)** Levato & Abt (1978); **j)** Abt & Morrell (1995); **k)** Hipparcos Catalog (ESA 1997); **l)** Holmberg, Nordstrom, & Andersen (2009); **m)** Grenier 1999 ; **n)** Roeser & Bastian 1988; **z)** This work

**Table 14**. Astrophysical data for the wide component of BU 627 AB

| Comp. | V | K | J-H | H-K | Sp | $\mu(\alpha)$ (mas yr$^{-1}$) | $\mu(\delta)$ (mas yr$^{-1}$) | Distance (pc) | Nature |
|---|---|---|---|---|---|---|---|---|---|
| D | 12.49 | 10.15 | +0.50 | +0.07 | K2V | -14.1 ± 2.7 | +10.6 ± 3.1 | 161 | optical |
| E | 12.61 | 8.79 | +0.76 | +0.15 | K8III | -0.3 ± 5.5 | -3.8 ± 9.2 | 3800 | optical |
| F | 14.13 | 12.75 | +0.18 | +0.09 | F7 | … | … | … | ¿? |

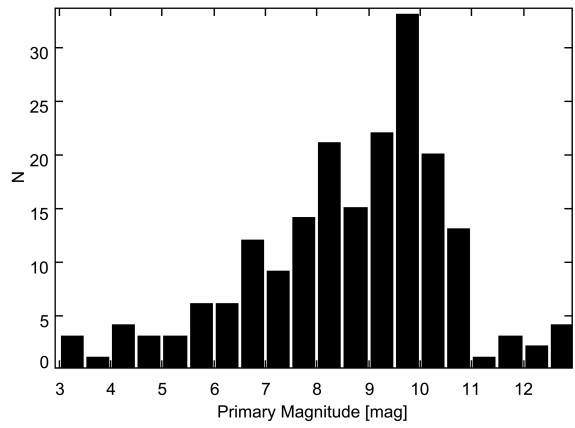

**Figure 1**. Magnitude distributions for the primary component.

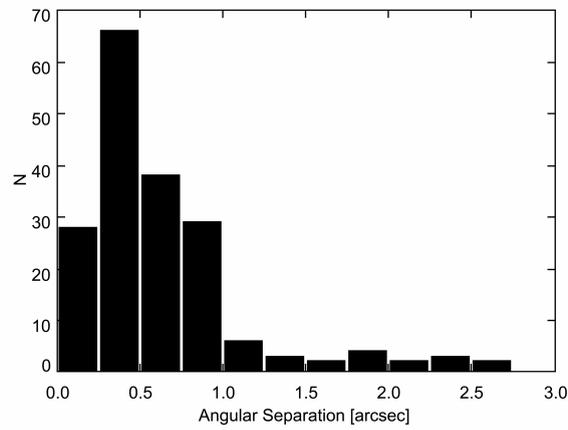

**Figure 2.** Distribution of angular separations (ρ), in arcseconds, for the binaries measured in this work.

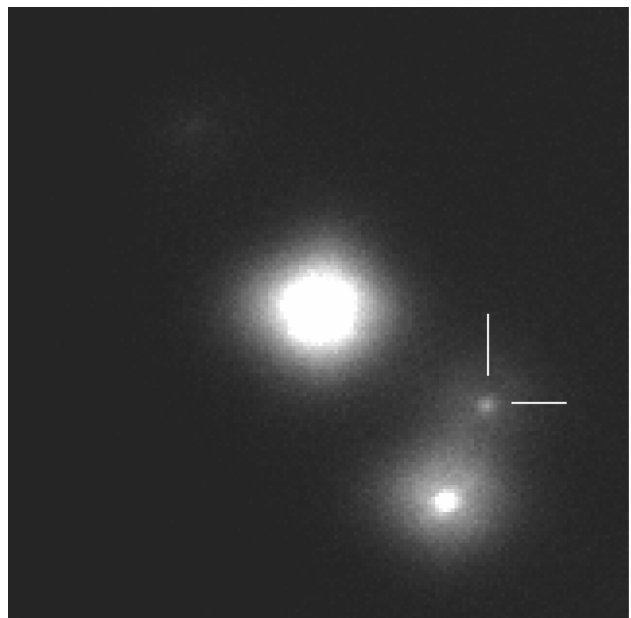

**Figure 3**. Binary star J 621 AB and the new companion.

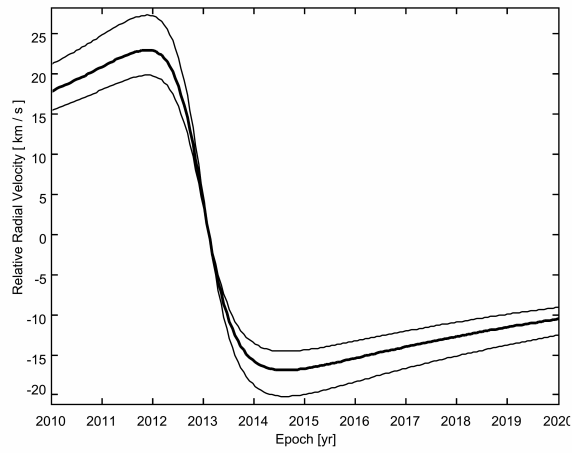

**Figure 4.** Curve of relative radial velocity (in km s$^{-1}$) for HU 718 in 2010-2020. The orbital parameters were used to calculate the radial velocities. Gray lines represent the Hipparcos parallax error.

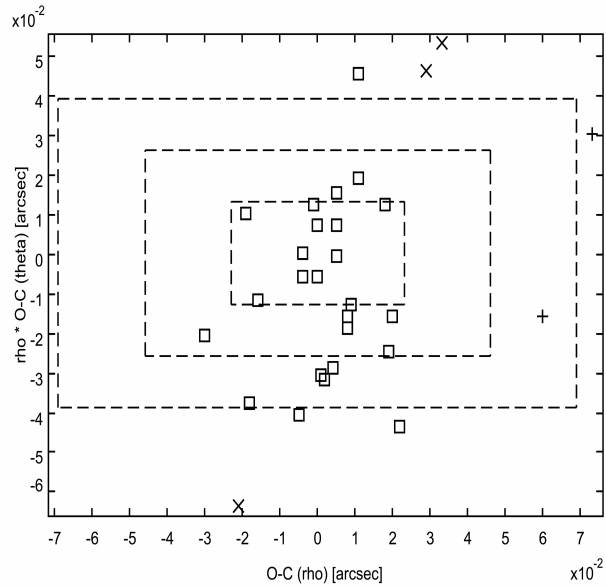

**Figure 5.** Residuals of the observational points for binaries with graded 1-2 orbits, in the tangential ($\rho \times \Delta\theta$) and radial ($\rho$) directions. This plot presents (O-C) differences between the observed and calculated positions. The central rectangles show $1\sigma$–$2\sigma$–$3\sigma$ bands in radial and tangential directions corresponding to the rms of O-C. "x" symbols represent binaries showing orbits with systematic runoff with respect to recent measures. "+" symbols are binaries with suspected runoff orbits.

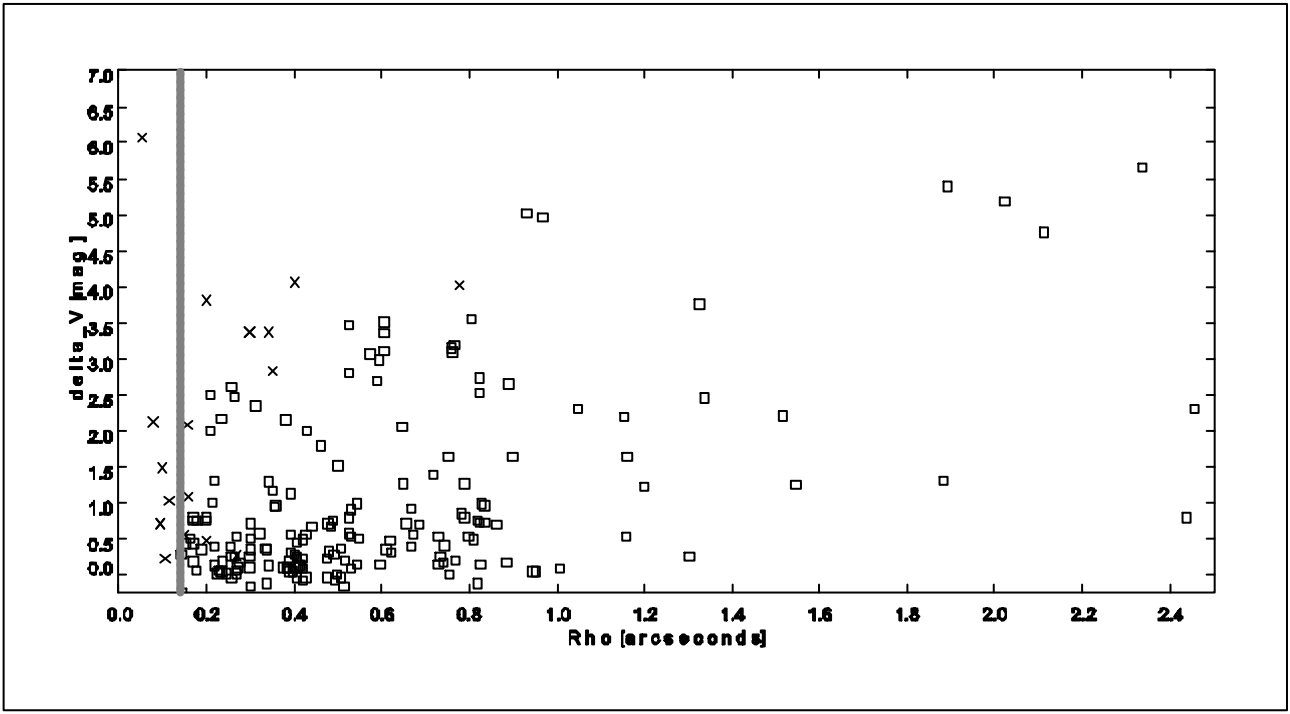

**Figure 6.** Positive and negative observations in the ρ–Δm space. Squares and crosses are positive and negative observations, respectively. The diffraction limit of the telescope is represented by a vertical line at ρ = 0.14".

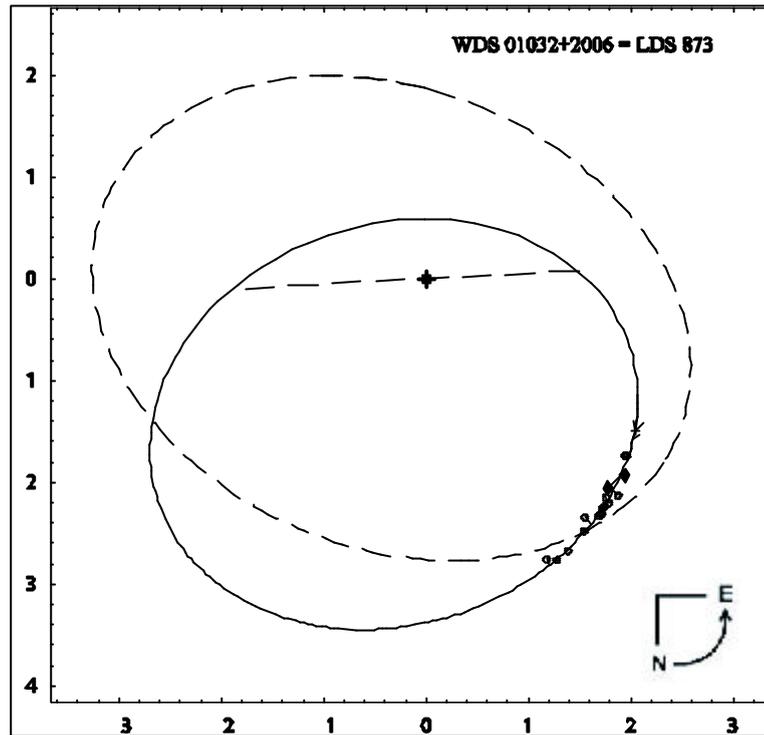

**Figure 7.** Apparent orbit for LDS 783. In this figure and the followed, the x and y scales are in arcseconds. The solid curve represents the newly determined orbit, while the dashed curve represents the previous orbit. The line passing through the origin indicates the line of nodes. Speckle measures are shown as filled squares, visual interferometric observations are shown as open circles, visual measures as plus signs, and measures from the ESA Hipparcos instrument are indicated as filled diamonds. The rejected observations are shown as crosses. All measures were connected to their predicted positions on the new orbit by O - C lines

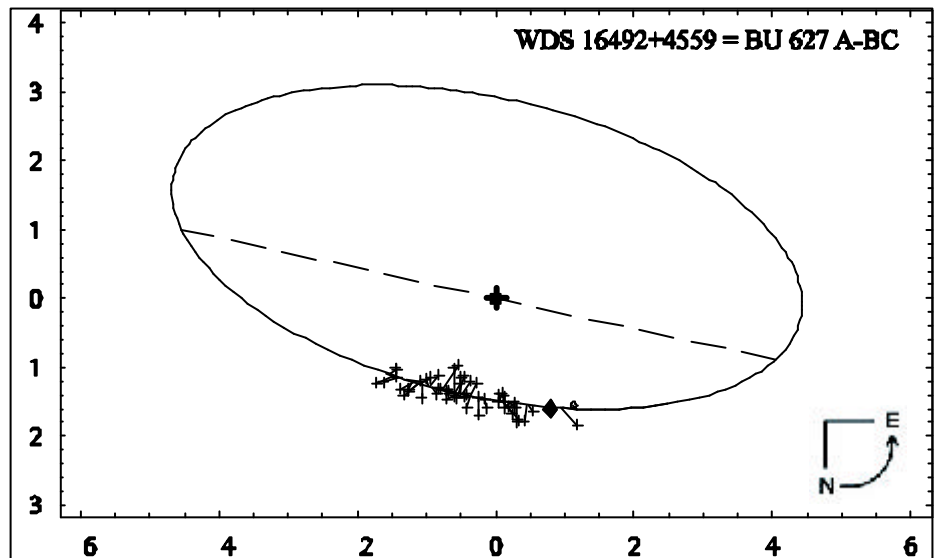

**Figure 8.** Apparent orbit for BU 627 A-BC. Symbols are the same as Figure 7

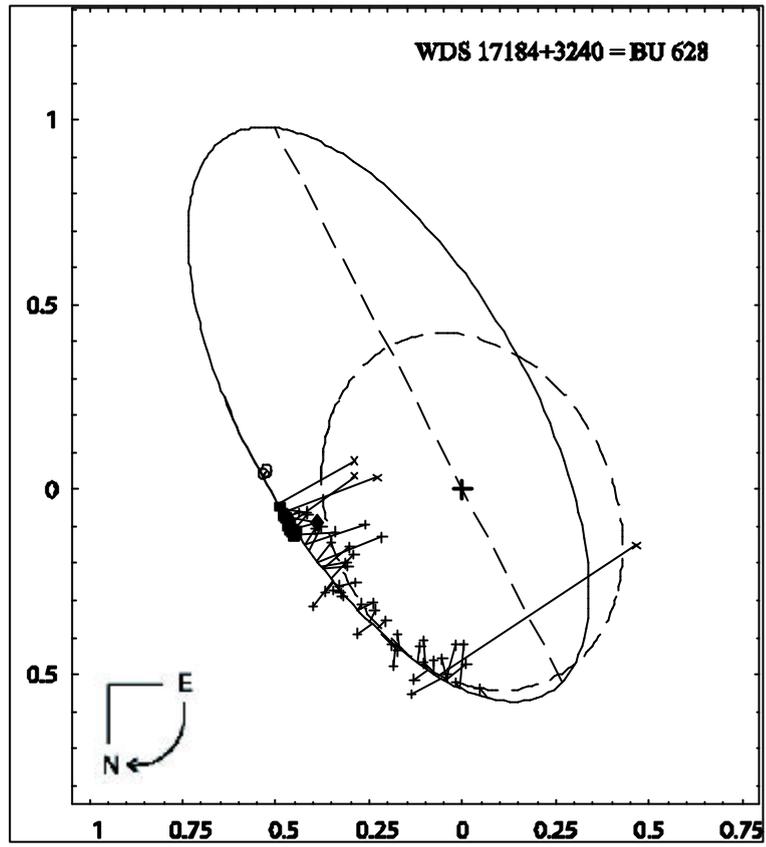
**Figure 9.** Apparent orbit for BU 628. Symbols are the same as Figure 7.

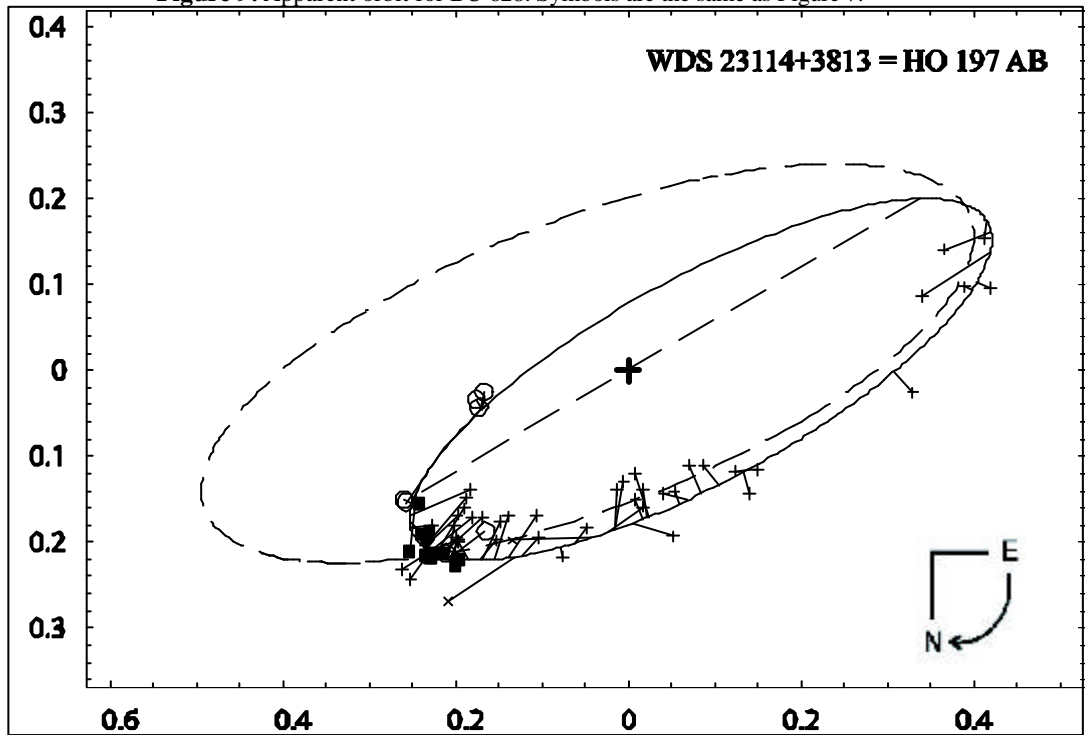
**Figure 10.** Apparent orbit for HO 197 AB. Symbols are the same as Figure 7.